\documentclass[
twocolumn,aps,prl,superscriptaddress
]{revtex4}

\usepackage{graphicx,color,hyperref}

\usepackage{amsmath}
\usepackage{amsfonts}
\usepackage{amssymb}

\usepackage{braket}
\usepackage[utf8]{inputenc}
\usepackage{amsmath}
\usepackage{bbold}

\usepackage{amsfonts}
\usepackage{amssymb}
\usepackage{graphicx}
\usepackage{braket}
\usepackage{float}
\usepackage{natbib}
\usepackage{hyperref}

\usepackage{tikz}
\usepackage{amsmath}

\newcommand{\tr}{\mathrm{tr}}
\newcommand{\str}{\mathrm{str}}

\usepackage[normalem]{ulem} 

\begin{document}

\title{
Theory of Anderson localization on the hyperbolic plane 
 }

\author{Alexander Altland}
\affiliation{Institut f\"ur Theoretische Physik, 
Universit\"at zu K\"oln, Z\"ulpicher Str. 77, 50937 Cologne, Germany
}
\author{ 
Tobias Micklitz 
}
\affiliation{Centro Brasileiro de Pesquisas F\'isicas, 
Rua Xavier Sigaud 150, 22290-180, Rio de Janeiro, Brazil 
} 
\author{Devasheesh Sharma}
\affiliation{Institut f\"ur Theoretische Physik, 
Universit\"at zu K\"oln, Z\"ulpicher Str. 77, 50937 Cologne, Germany
}
\author{Maksimilian Usoltcev}
\affiliation{Institut f\"ur Theoretische Physik, 
Universit\"at zu K\"oln, Z\"ulpicher Str. 77, 50937 Cologne, Germany
}
\author{Carolin Wille}
\affiliation{ London Centre for Nanotechnology, University College London, London WC1H 0AH, United Kingdom}
\date{April 27, 2026}

\begin{abstract}
The two-dimensional hyperbolic plane, $\mathbb{H}^2$, is an unusual system in
that dimensionality changes with scale: locally two-dimensional and planar at short distances, but effectively infinite-dimensional at large scales, it provides an
interesting paradigm for the study of (quantum) phase transitions, notably the
disorder-driven Anderson transition. Generalizing previous work, which treated
short and large distance scales separately, we develop a unified framework interpolating between the principles
of low- and high-dimensional Anderson localization. As a main result, we derive a
two-parameter flow in a plane spanned by scale-dependent curvature (setting the
system's effective dimensionality) and
conductivity, with an extended critical line separating metallic and insulating
phases.
\end{abstract}

\maketitle

{\it Introduction:---} Physical dimensionality is one of the key parameters
governing the long-distance and critical behavior of complex systems. Unlike
symmetries or emergent collective phenomena, it is typically treated as an
immutable attribute, organizing criticality into well-defined universality
classes. The hyperbolic plane, $\mathbb{H}^2$, provides a notable exception to
this paradigm~\cite{Wang2025BoundarySensitive,Okunishi2024BoundaryCorrelations,
Mosko2025VertexHyperbolicTN,Banerjee2026BetheCritical,
HyperbolicPercolation2025,Boettcher2015,Li2024,Chen2024,Curtis2025}. As a
two-dimensional surface of constant negative curvature with radius $L$, it
exhibits a scale-dependent effective dimensionality: at short distances, $r \ll
L$, the geometry is locally flat with $d_{\mathrm{eff}}=2$, while at large
distances, $r \gg L$, exponential volume growth renders the space effectively
infinite-dimensional, $d_{\mathrm{eff}} \to \infty$. In this way, $\mathbb{H}^2$
provides a setting in which regimes associated with different dimensionalities
coexist.

This crossover is clearly reflected in quantum transport, where, depending on
spatial dimension $d$, the localization of quantum states by disorder is
governed by different principles. For $d \le
2$, the high return probability of semiclassical
trajectories leads to constructive interference and ---  in generic,
non-interacting systems --- to localization for arbitrarily weak disorder. For
$d > 2$, returns become irrelevant, and localization requires 
strong disorder to overcome escape into the exponentially large volume.

\begin{figure}[t!]
    \centering
    \vspace{.1cm}
    \includegraphics[width=6.5cm]{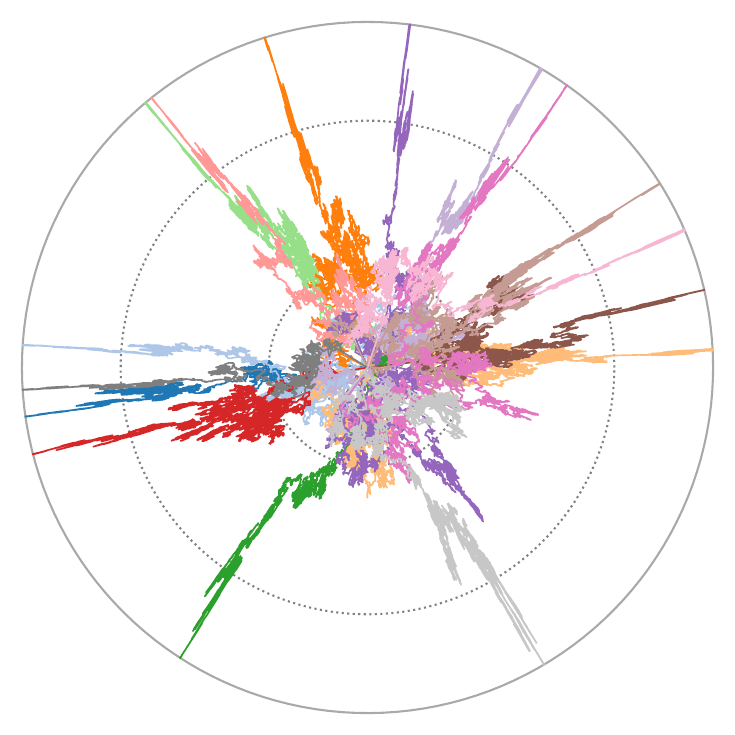}
    \vspace{-.2cm}
    \caption{A number of diffusive Brownian motion processes simulated on
    $\mathbb{H}^2$. The dashed curves indicate geodesic circles of radii $r=(1,2,5) L$.
    Note the crossover from diffusive to ballistic behavior, as trajectories
    escape to infinity. 
 }    
\label{fig:figure1}
\end{figure}

While conventional systems realize either one of these regimes, the hyperbolic
plane exhibits both within a single setting. Previous work has addressed these
limits separately, using weak-disorder perturbation theory~\cite{Curtis2025} and
lattice exact diagonalization~\cite{Li2024,Chen2024}. The aim of the present
work is to develop a unified theory
 and to analyze the resulting dimensional crossover. Central to this construction is a
scale-dependent change in the semiclassical point-to-point transmission
probability --- from two-dimensional diffusive behavior at short distances to
effectively ballistic propagation at large scales (see
Fig.~\ref{fig:figure1}) --- which mediates an interpolation between the two
localization mechanisms.

We will address this problem using a renormalization group approach
generalized to curved backgrounds, complemented by a transfer-matrix analysis
for lattice discretizations of $\mathbb{H}^2$. The combination of these concepts
will lead to
a two-parameter flow in a phase plane spanned by curvature and conductivity,
with an extended critical line separating metallic and insulating regimes, the
main result of this work.

{\it Definition of the problem:---} 
The hyperbolic plane may be realized as the upper sheet of the
two-sheeted hyperboloid $x_1^2+x_2^2-x_3^2=-L^2$, $x_3>0$,
in ambient Minkowski space $\mathbb R^{2,1}$ with metric
$ds^2_{\rm amb}=dx_1^2+dx_2^2-dx_3^2$. The radius $L$ determines the  negative scalar curvature, $R = -2L^{-2}$.  
In  polar coordinates, $x=(r,\phi)$ the metric tensor takes the form  
$ds^2 = dr^2 + L^2 \sinh^2\left(r/L\right) d\phi^2$,  
where $\phi \in [0,2\pi]$ is an angle and $r$ the geodesic distance from the origin.

We are interested in the physics of quantum particles of characteristic energy
$E=k_0^2/2m$ governed by the random Hamiltonian
$H = -\frac{1}{2m}\Delta + V(x)$. Here,  
$\Delta = g^{-1/2}\partial_\mu\!\left(\sqrt{g}\,g^{\mu\nu}\partial_\nu\right)$
is the Laplace operator on the plane, which in the chosen coordinates assumes the form
\begin{align}
   \Delta  
   &=
   \partial_r^2 +\coth \left( \frac{r}{L} \right) \frac{1}{L} \partial_r 
   + 
   \frac{1}{L^2 \sinh^2 \left( \frac{r}{L} \right) } \partial_\phi^2,
\end{align} 
with $g=\det(g_{\mu \nu})=L^2 \sinh^2(r/L)$.
We assume the random potential $V$ to
be  $\delta$-correlated  with a 
variance $\left\langle V(x)V(y) \right\rangle = \left(
W/k_0  \right)^2 |g(x)|^{-1/2}
\delta(x-y)
$ 
setting the disorder strength, $W$.

Before turning to the quantum dynamics of this problem, consider the classical
limit, whose dynamics is governed by the `heat operator'  $(\partial_t - D \Delta)$
with a constant $D$ (we hesitate to call it diffusion
operator, as the dynamics is diffusive only locally). The   matrix elements $\Delta^{-1}(x,y)\equiv
\braket{x|\Delta^{-1}|y}$ are  measures for (the negative of) the time integrated probability of a
particle to propagate from $y$ to $x$, and they are known in terms of
eigenfunction decompositions  (cf. the Supplementary Material) with the
simple result $\Delta^{-1}(x,y)=\frac{1}{2 \pi}\ln \tanh(d(x,y)/2L)$, where
$d(x,y)$ is the geodesic distance between the observation points. Specifically,
for  $d(x,y)= r\ll L$ small compared to the curvature, $\Delta^{-1}(r)\approx
\frac{1}{2\pi}\ln(r/2L)$ as for a diffusive Euclidean plane. Conversely, for
$r\gg L$, $\Delta^{-1}(r)\approx - \frac{1}{\pi}\exp(-r/L)$. This exponential
suppression indicates directed isotropic decay towards any of the points on the
circle of radius $r$, and metric circumference $\sim \exp(r/L)$. Indeed, the
analysis of the full space-time dependent heat equation shows that in this
regime $\langle r \rangle\sim 2Dt/L$ grows ballistically, consistent with the
escape dynamics of the trajectory simulations in Fig.~\ref{fig:figure1}.

{\it Effective field theory:---}Turning to the quantitative analysis,
we consider the uniquely determined extension of the effective field theory of
Anderson localization~\cite{Efetov1997,Altland2023} to a background geometry
with a Riemannian metric~\cite{repo}, $g_{\mu\nu}$. This is a nonlinear $\sigma$-model defined by the
functional integral $Z = \int DQ\, e^{-S[Q]}$, with the
action~\cite{usoltcev2026continuumfieldtheorymatchgate,repo}, 
\begin{align}
\label{eq:nlsm}
S[Q] 
&= 
-\frac{\sigma}{16} 
\int d^2x\sqrt{g}\, \mathrm{str} \left( Q \Delta Q \right).
\end{align}
Here, $Q=T \tau_3 T^{-1}=\{Q^{\alpha \beta}(x)\}$ where the index
$\alpha=(a,c,r)$ comprises  an advanced/retarded index $a=\pm 1$ (with 
$\tau_3$
being a Pauli matrix in this space), an index $c=\pm 1$ for time-reversal, and a
`super-index' $r=\mathrm{b,f} $ distinguishing between commuting ($\mathrm{bb,ff} $) and
anti-commuting ($\mathrm{bf,fb} $) matrix entries, with the supertrace 
$\str(A)=\tr(A^\mathrm{bb})-\tr(A^\mathrm{ff})$. (Anticipating the need for
non-perturbative constructions in the strong-disorder regime, we here work with the
supersymmetry formalism, where $Q$ are $8\times 8$ supermatrices, 
time-reversal symmetry (symmetry class AI) requires~\cite{Efetov1997}  
$T$ to take values in the ortho-symplectic supergroup $\mathrm{OSp}(2,2|4)$, and
$Q $ in a coset space $\mathrm{OSp}(2,2|4)/\mathrm{OSp}(2|2)\times \mathrm{OSp}(2|2)$
dividing out transformations commuting with $\tau_3$.)
Finally, the integral is over the
invariant area element, defined by $ \sqrt{g}dr d\phi=L \sinh(r/L)dr d\phi$,
approaching the planar element $rdrd\phi$ for $r\ll L$ and exponentially growing
otherwise. 


{\it Renormalization:---}$\sigma$-models in two-dimensional space generically
fluctuate strongly at large length scales, leading to an effective downward
renormalization of the coupling constant, and eventually
localization~\cite{EversMirlin2008}. Our goal is to find out how the
geometric distortion, formally encoded in the volume element and the Laplacian,
interferes with this picture. In concrete terms, this requires an RG protocol,
implemented, e.g., by mode renormalization: depending on their Laplace
eigenvalues, one organizes the field space $T\to T_\textrm{f} T_\textrm{s}$ into
`fast' and `slow' modes, and proceeds by iterative integration over the
fast $T_\textrm{f}$. 

While in translationally invariant settings the corresponding eigenfunctions are
just Fourier modes, we here need to work with the  eigenfunctions of the
hyperbolic Laplacian. These functions provide a convenient, physical basis
for the expansion of field fluctuations in   a variant of the Fourier transform
known as the Fourier-Helgason transform. Apart from this modification, the
one-loop integration over fast field modes with eigenvalues in an interval
$[\Lambda e^{-\ell},\Lambda]$ below a UV cutoff is standard~\cite{Efetov1997,Altland2023} (see the
Supplementary Material for details), and it leads to the coupled flow
equations
\begin{align}
    \label{eq:TwoParameterFlow}
     \frac{ d \sigma}{d \ell}&= -\frac{1}{2\pi} \frac{u^2\tanh(\pi
     u)}{u^2+\frac{1}{4} }  ,\nonumber \\
     \frac{d u}{d \ell}& = - u,
\end{align}
where $(\sigma,u)=(\sigma(\ell),u(\ell))$, are scale dependent conductivity and
curvature radius, respectively, with initial data  given by
$(\sigma(0),u(0))=(\sigma_0,\Lambda L)$. 

To make sense of these equations, note that in the limit of infinite curvature
radius, $u\to \infty$, the first, $d_\ell \sigma \approx
-1/2\pi$, describes the standard downward renormalization  of the conductivity as
$\sigma(\ell)=\sigma_0- \frac{\ell}{2\pi} $ by weak localization. More
generally, the solution of Eq.~\eqref{eq:TwoParameterFlow}
yields $u(\ell)=\Lambda L\exp(-\ell)$, and 
\begin{align}
    \label{eq:SigmaOfb}
    \sigma(\ell)=\sigma_0 - \frac{1}{2\pi} \int^{\Lambda L}_{\Lambda  Le^{-\ell}} 
    \frac{d\lambda \lambda \tanh(\pi \lambda)}{\lambda^2+1/4}.
\end{align}
Physically, the integral on the right extends over hyperbolic diffusion modes
with eigenvalues  $\mathrm{EV}( \Delta) =-L^{-2}(\lambda^2+1/4)$, where
$\lambda$ assumes the role of a scalar momentum, and the `density of modes' (aka
 Fourier-Plancherel measure) $d \lambda \lambda \tanh(\pi \lambda)/\pi
\stackrel{\lambda \to \infty}\longrightarrow  \lambda d\lambda/\pi  $ approaches the radial
measure of the plane at short distance scales or large momentum. In the opposite
limit, a conspiracy of a low mode density and a spectral gap $1/4$
suppresses the effect. (Note that the same mechanism is responsible for the crossover from diffusive to effective escape 
dynamics 
shown by the simulations of Fig.~\ref{fig:figure1}.)

\begin{figure}[h]
    \centering
    \vspace{.1cm}
    \includegraphics[width=7.5cm]{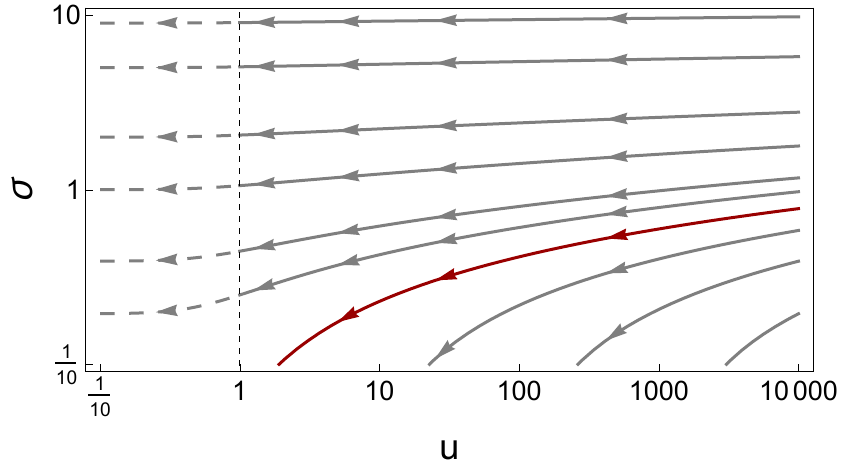}
    \vspace{-.2cm}
    \caption{Flow of the two parameters $(\sigma,u)$ as a function of the RG
    length scale. A separatrix (red line) divides the parameter plane into a
    metallic and an insulating regime. In the metallic regime (above the
    separatrix), the RG flow terminates ($u(\ell) \sim 1$) at finite, non-zero
    conductivity, while in the complementary, localizing, regime, the
    conductivity flows to zero on an RG length scale below the curvature scale
    and localization occurs before the effect of the curvature is felt. }   
\label{fig:figure2}
\end{figure}

Fig.~\ref{fig:figure2} visualizes the flow $(\sigma,u)(\ell)$ traced out by
the solutions above. Note the extreme efficiency of the  hyperbolic plane in
protecting metallicity: for systems with weak to moderate disorder $\sigma>1$,
even a small curvature with large radius $L \sim \Lambda^{-1} \exp(2\pi \sigma)$
suffices to block strong localization. In this parameter domain, the
cutoff $u(\ell)\sim 1$ below which loop interference becomes ineffective is
reached at finite residual values of the conductivity. The boundary to the
localized regime is marked by a separatrix, $\sigma_\mathrm{sep}(u) $ defined by
the condition that $\sigma_\mathrm{sep}(u\to 1) \lesssim 1 $, i.e. a
conductivity suppressed down to small values at $u\sim 1$. Finally, note that the $\beta$-function
defined by the right-hand side of Eq.~\eqref{eq:TwoParameterFlow} stays
negative for all values of its argument: loop interference, even if weak, is a
uniformly suppressing mechanism. As a consequence, our analysis so far predicts
a crossover  between
metallic and insulating behavior, but no phase transition. However, this is not
the end of the story: for $u \lesssim 1$, i.e. short distance cutoff comparable
to curvature radius,   the principles of loop
localization give way to those of effectively high-dimensional systems. To
describe the impact of these, we abandon the continuum theory and discretize
into a lattice representation with spacing $\Lambda^{-1} \exp(\ell)\equiv 1
\approx L$.

\begin{figure}
    \centering
    \vspace{.1cm}
    \includegraphics[width=8.5cm]{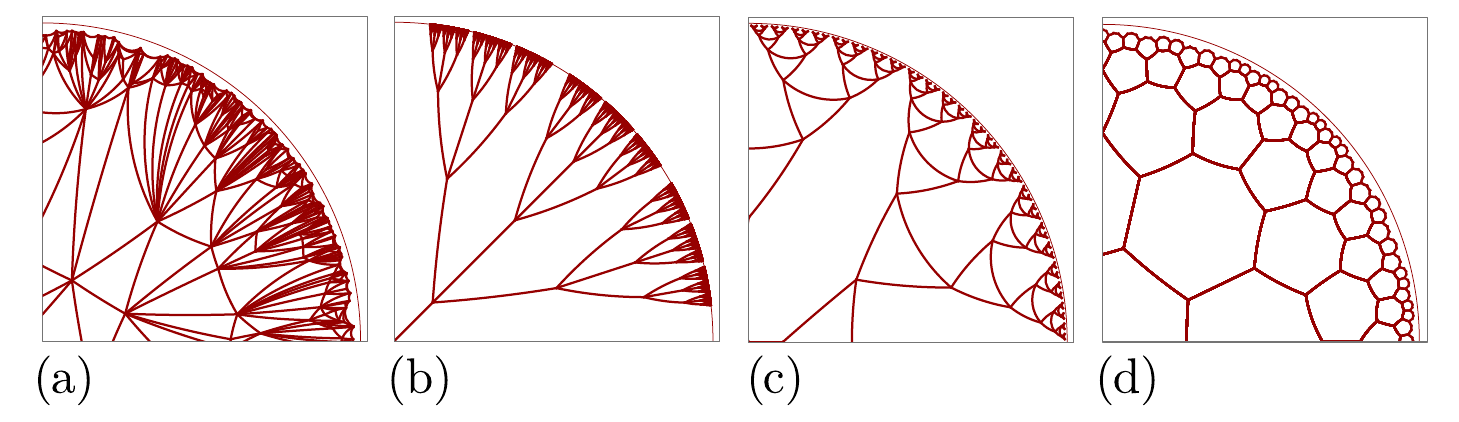}
    \vspace{-.2cm}
    \caption{Discretizations of the hyperbolic plane: (a) Poisson–Delaunay
    random graph, (b) $q=4$ Bethe lattice, (c) $(p,q)=(3,2)$
    Husimi-cactus graph, (d) $\{p,q\}=\{7,3\}$ regular
    polygon tessellation. Discussion, see text.   
 }    
\label{fig:figure3}
\end{figure}

{\it Metal-insulator transition:---}The discretization of the action 
Eq.~\eqref{eq:nlsm} reads
\begin{align}
    \label{eq:lattice_nlsm}
    S[Q] = \sum_{\langle ij\rangle}w_{ij}  \mathrm{str}(Q_i  Q_j),    
\end{align}
where the  set of points $\{i\}$ and weights $\{w_{ij}\}$ approximates the
 continuum geometry. Referring to the Supplementary Material for details, there
 are various different discretizations to choose from, including
 (cf. Fig.~\ref{fig:figure3}), (a) the random Poisson-Delaunay
 (PD)\cite{BenjaminiPaquettePfeffer2014}   discretizations with weights $w_{ij}$
 defined by matching to the continuum Laplacian, (b) Bethe lattices with
 coordination number $q$, (c) Husimi $(p,q)$-cactus trees, with $q$
 $p$-gons meeting at vertices, and (d) regular $\{p,q\}$ polygon tessellations.


Referring to the Supplementary Material for a short comparison, the suitability
of specific choices (a-d) as  proxies of $\mathbb{H}^2_L$ depends on three
criteria: the most obvious one, exponential growth in metric area is met by all
realizations (a-d). (In Fig.~\ref{fig:figure3}, the geodesic point-to-point
distance is scaled to a uniform value, i.e. the increasing point density at the
boundary reflects area growth.) The second is the loop content of the lattice,
where `natural' tessellations of $\mathbb{H}^2$ in terms of local polygon structures
(a,c,d) differ from the loopless Bethe lattice (b).  Finally, we require optimal
matching of the lattice and the continuum  Laplacian at scales $\sim L$, a
condition built into the PD discretization (a) by design. While previous
numerical work  has focused on
(d)~\cite{Chen2024,Li2024}, the matching of all three criteria  makes
(a) our model of choice.

{\it Lattice localization:---} Referring to the
Supplemental Material for a short review, classic
work~\cite{Efetov1997,ZIRNBAUER1986375} on lattice-localization described
localization on 
Bethe lattices with uniform bond weights $w_{ij}=w$ in terms
of  recursion relations 
\begin{align}
    \label{eq:BetheTransfer}
    \Psi(Q)=\int dQ'\, e^{-w\,\str(QQ')}\,\Psi^K(Q')
\end{align} 
 where $\Psi(Q)$ is the functional integral over all children  connected to
a given root with field amplitude $Q$, recursively represented in terms of the
first $K=q-1$ daughter sites. 

Deep in the localized regime, $w\ll 1$,
these integrals are dominated by the leading noncompact variables $t$
parameterizing the matrix manifold, $Q\to Q(t)$. Supersymmetry implies that
the equation is trivially solved by  $\Psi(Q)\equiv \Psi(t)=1$, and the defining
feature of the localized phase is that this solution is unique~\cite{Efetov1997}.
The  metal-insulator transition is then diagnosed by the emergence of a second
solution, detected by stability
analysis  of the linearized form $\Psi=1+\Phi$ in terms of the integral kernel
defined through Eq.~\eqref{eq:BetheTransfer}. As a result, one finds that the
transition occurs at a critical coupling $w^* \sim 1/(K\ln K)^2\ll 1$,  
corresponding to a disorder strength $W^*\sim w^{* -1/2}\sim K \ln K$. To
understand this result, note that in a theory with unit hopping strength, and
disorder amplitude
$W\lesssim K$,
statistically at least one of $K$ nearest neighbor
sites will be resonantly coupled to the root site, and hence be able to support
a quantum state hybridized with it. The logarithmic correction in the criterion
results from the extension of the argument to higher order site generations. 

The description of localization on the PD lattice requires a generalization of
the construction to the presence of loops and bond-strength inhomogeneities.
Concerning the former, in the Supplementary Material we consider the Husimi
cactus graph as an analytically tractable proxy for general hyperbolic lattices
in which loops are present, but do not form extended networks. This class of
graphs captures the minimal extension beyond tree-like structures and is
sufficient in settings where repeated loop excursions are statistically
suppressed. We derive and solve an exact generalization of
Eq.~\eqref{eq:BetheTransfer}, with the result that, perhaps not surprisingly,
short loops on a lattice of strong negative curvature do not qualitatively
change the localization criterion: quantum states are dominantly correlated
along escape paths that avoid self-returning excursions. (This finding is
consistent with a recent analysis of Husimi-cactus graphs~\cite{bandeira2026} in
terms of cavity Green-function methods.)

A second deviation of the PD discretization from the uniform Bethe lattice is
the presence of bond-strength inhomogeneities of $\mathcal{O}(1)$ around an
average value $w=\left\langle w_{ij} \right\rangle$. In the Supplementary
Material we demonstrate that this feature, too, does not significantly change
the transition criterion; localization still requires disorder strength $W^\ast
\sim K$, proportional to the number of bonds along which wavefunctions can
hybridize. Finally, while we did not analyze the combined effect of loops and
bond statistics, we believe that the overall picture remains the same. As a
corollary, the critical exponents characterizing the transition are expected to
coincide with the mean-field-like~\cite{EversMirlin2008} Bethe-lattice
exponents: $\nu=1$ for the correlation-length exponent, and $\beta=1$ for the
local-density-of-states/order-parameter exponent.

{\it Combined theory:---} Having discussed both weak and strong localization in
the respective limits, we may now combine them to a unified framework. This
discussion is
organized around starting configurations $(\sigma_0,u_0)$ in the vicinity of
the separatrix, i.e. $\sigma_0\sim \ln(u_0)$. For these values, the RG flow
sends us to $(\sigma(\ell),u(\ell))=\mathcal{O}(1,1)$, conductivity values of
$\mathcal{O}(1)$ when the curvature has become comparable to the UV cutoff. PD
discretization then maps the system onto one with bond strength $\langle w_{ij}
\rangle=\mathcal{O}(1)$ averaging to a unit value, parametrically consistent
with the coupling $w^\ast=\mathcal{O}(1)$ of the lattice
\emph{transition}. In this way, the crossover separatrix of the continuum theory
gets promoted to a true critical line of the combined theory. As is often the case with
theories of strong localization, the latter operates in an $\mathcal{O}(1)$
corridor of the coupling constants, lacking control by a small parameter.
(However, some of these $\mathcal{O}(1)$ matches realize nontrivial consistency
conditions: for example, the PD discretization with its optimal matching to the
continuum Laplacian is required to avoid $q$-dependent non-universal predictions
of localization thresholds that would be obtained for competing discretizations
in terms of regular $\{p,q\}$ polygon tessellations or cactus graphs.)

We finally note that the scale dependent resist\emph{ivity}, $ \rho=\sigma^{-1}$
entering as a parameter of the critical flow, must be distinguished from the
resist\emph{ance} that would be measured in actual transport probes. To
appreciate this point, consider a `two-terminal' setup reflecting the system's
symmetry, e.g. the current $I$ flowing through a ring of radius $d$ at bias
voltage $V$ centered around an origin connected to a grounded electrode of small
radius $\epsilon$. Obtaining the scale dependent resistance as $R(d)=V/I$
requires integrating the continuity equation, or, closer to the present
framework, analyzing the action cost of a voltage gradient. In the Supplementary
Material we show that this yields the resistance as
\begin{align}
    \label{eq:ConductanceFromSigma}
    R(d) 
=    \frac{\rho(d)}{ 2\pi}\ln\! \left( \frac{\tanh(d/2L)}{\tanh(\epsilon/2L)} \right).   
\end{align}
For small system sizes, $d\ll L$, $ R(d)\simeq \frac{\rho(d)}{2\pi}
 \ln(d/\epsilon)$ reduces to a standard result for the resistance of an annulus
 geometry, i.e. resistance=resistivity $\times$ a logarithmic  geometric
 factor~\footnote{As a sanity check, for a thin annulus $d= \epsilon + \delta
 \approx \epsilon$, this reduces further to the sheet resistance, $R\approx
 \frac{\rho \delta}{2\pi \epsilon}  $, with a geometry factor equal to the ratio
 of sample thickness/width.}. However, for $d\gg L$, the geometric
 $d$-dependence levels off at the value $R(d) \to \frac{\rho(L)}{2\pi}
 \ln(2L/\epsilon) $, where we took into account that loop renormalization ceases
 to be effective in this regime. In other words, the two-terminal resistance of
 arbitrarily large geometries is entirely dominated by the resistivity of
 regions $\lesssim L$ near the center electrode. This result reflects the
 exponential area increase at $d\gtrsim L$, implying that for finite
 $\sigma(L)>1$ (metallic regime) the outer regions of the ring act like a giant
 parallel shunt which does not add to the series resistance of the full system. 

{\it Discussion:---} The effective dimensionality of the hyperbolic plane
changes with the observation scale, from  two-dimensional at the smallest
distance scales to $d=\infty$ at the largest. We here explored how this
transmutation manifests itself in terms of an effective theory 
integrating localization by loop interference in the  low-, and
by strong disorder in the high-dimensional limit, respectively. The two pictures match
in the crossover region, leading to  the prediction of 
a global critical line in the (conductivity, curvature) parameter plane. This  
unified picture is consistent with previous analytical~\cite{Curtis2025} and 
numerical work~\cite{Chen2024,Li2024} on the
former and latter regime, respectively.

The structures discussed here may have ramifications  in the context of low-dimensional holography. For example, the $\sigma$-model on $\mathbb{H}^2$ --- albeit
in its class~D incarnation --- is an effective model for random matchgate tensor
networks~\cite{usoltcev2026continuumfieldtheorymatchgate}, which in turn have
been discussed~\cite{Jahn2019MatchgateHolography,Jahn2022Boundary} as proxies
for constant time slices in AdS$_3$ geometries, or Euclidean AdS$_2$. In both
cases, the ensemble perspective is crucial for introducing boundary
correlations, and it will be interesting to explore if the structures
discussed here can be of relevance in these contexts.

{\it Acknowledgements:---} M.U. thanks Julian Arenz for fruitful discussions.
This work was supported by the Deutsche
Forschungsgemeinschaft (DFG, German Research Foundation) within the CRC network TR 183
(project grant 277101999) as part of projects A03, the Brazilian funding agencies
CNPq and FAPERJ, and by the EPSRC UK Quantum Technologies Programme under grant EP/Z53318X/1.

\bibliography{bibliography}

@misc{usoltcev2026continuumfieldtheorymatchgate,
      title={Continuum field theory of matchgate tensor network ensembles}, 
      author={Maksimilian Usoltcev and Carolin Wille and Jens Eisert and Alexander Altland},
      year={2026},
      eprint={2603.06202},
      archivePrefix={arXiv},
      primaryClass={cond-mat.dis-nn},
      doi = {10.48550/arXiv.2603.06202},
      url={https://arxiv.org/abs/2603.06202}, 
}

@article{Hashimoto1989,
  author  = {{Ki-ichiro} Hashimoto},
  title   = {Zeta functions of finite graphs and representations of p-adic groups},
  journal = {Adv. Stud. Pure Math.},
  volume  = {15},
  year    = {1989},
  pages   = {211--280}
}

@book{Terras2010,
  author    = {Audrey Terras},
  title     = {Zeta Functions of Graphs},
  publisher = {Cambridge University Press},
  year      = {2010}
}

@article{BenjaminiPaquettePfeffer2014,
  author  = {Itai Benjamini and Elliot Paquette and Joshua Pfeffer},
  title   = {Anchored expansion, speed, and the hyperbolic Poisson Voronoi tessellation},
  journal = {Annals of Probability},
  year    = {2017},
  eprint  = {arXiv:1409.4312}
}

@misc{bandeira2026,
      title={Anderson Localization on Husimi Trees and its implications for Many-Body localization}, 
      author={Dafne Prado Bandeira and Marco Tarzia},
      year={2026},
      eprint={2601.04155},
      archivePrefix={arXiv},
      primaryClass={cond-mat.dis-nn},
      url={https://arxiv.org/abs/2601.04155}, 
}

@article{ZIRNBAUER1986375,
title = {Anderson localization and non-linear sigma model with graded symmetry},
journal = {Nuclear Physics B},
volume = {265},
number = {2},
pages = {375-408},
year = {1986},
issn = {0550-3213},
doi = {https://doi.org/10.1016/0550-3213(86)90316-0},
url = {https://www.sciencedirect.com/science/article/pii/0550321386903160},
author = {Martin R. Zirnbauer}
}

@article{NielsenNock2009,
  author  = {Frank Nielsen and Richard Nock},
  title   = {Hyperbolic Voronoi Diagrams Made Easy},
  journal = {International Journal of Computational Geometry and Applications},
  year    = {2010},
  volume  = {20},
  number  = {3},
  pages   = {287--302},
  doi     = {10.1142/S0218195910003291},
  eprint  = {arXiv:0903.3287}
}

@article{Wang2025BoundarySensitive,
  author  = {Wang, X. and Zhang, Y. and Zhou, Y.},
  title   = {Emergence of a Boundary-Sensitive Phase in Hyperbolic Ising Models},
  journal = {Physical Review Letters},
  year    = {2025},
  note    = {in press}
}

@article{Okunishi2024BoundaryCorrelations,
  author  = {Okunishi, K. and Nishino, T.},
  title   = {Boundary Correlation Functions of the Ising Model on a Hyperbolic Lattice},
  journal = {Progress of Theoretical and Experimental Physics},
  volume  = {2024},
  pages   = {093A02},
  year    = {2024}
}

@article{Mosko2025VertexHyperbolicTN,
  author  = {Mosko, M. and Jahn, A. and B{\"u}hler, C.},
  title   = {Vertex Representation of Tensor Networks on Hyperbolic Lattices},
  journal = {Physical Review E},
  volume  = {111},
  pages   = {024105},
  year    = {2025}
}

@article{Banerjee2026BetheCritical,
	title={Critical Phenomena on the Bethe Lattice}, 
      author={Rudrajit Banerjee and Nicolas Delporte and Saswato Sen and Reiko Toriumi},
      year={2026},
      eprint={2601.01961},
      archivePrefix={arXiv},
      primaryClass={hep-th},
      url={https://arxiv.org/abs/2601.01961}, 
}

@article{HyperbolicPercolation2025,
  author  = {Curien, N. and Miermont, G.},
  title   = {Scaling Limits of Critical Percolation on Hyperbolic Random Triangulations},
  journal = {Random Structures \& Algorithms},
  year    = {2025},
  note    = {to appear}
}

@article{Li2024,
  author       = {Tianyu Li and Yi Peng and Yucheng Wang and Haiping Hu},
  title        = {Anderson transition and mobility edges on hyperbolic lattices with randomly connected boundaries},
  journal      = {Communications Physics},
  volume       = {7},
  pages        = {371},
  year         = {2024},
  doi          = {10.1038/s42005-024-01848-7}
}

@article{Osborn1991,
  author  = {Osborn, Hugh},
  title   = {Weyl consistency conditions and a local renormalization group equation for general renormalizable field theories},
  journal = {Nucl. Phys. B},
  volume  = {363},
  pages   = {486--526},
  year    = {1991}
}

@article{JackOsborn1990,
  author  = {Jack, I. and Osborn, H.},
  title   = {Analogs for the $c$ theorem for four-dimensional renormalizable field theories},
  journal = {Nucl. Phys. B},
  volume  = {343},
  pages   = {647--688},
  year    = {1990}
}

@article{Vassilevich2003,
  author  = {Vassilevich, Dmitri V.},
  title   = {Heat kernel expansion: user's manual},
  journal = {Phys. Rept.},
  volume  = {388},
  pages   = {279--360},
  year    = {2003},
  archivePrefix = {arXiv},
  eprint  = {hep-th/0306138}
}

@article{Chen2024,
  author  = {Anffany Chen and Joseph Maciejko and Igor Boettcher},
  title   = {Anderson localization transition in disordered hyperbolic lattices},
  journal = {Physical Review Letters},
  volume  = {133},
  number  = {6},
  pages   = {066101},
  year    = {2024},
  doi     = {10.1103/PhysRevLett.133.066101},
  eprint  = {arXiv:2310.07978},
  archivePrefix = {arXiv},
  primaryClass = {cond-mat.dis-nn}
}

@article{Curtis2025,
  author  = {Jonathan B. Curtis and Prineha Narang and Victor Galitski},
  title   = {Absence of Weak Localization on Negative Curvature Surfaces},
  journal = {Physical Review Letters},
  volume  = {134},
  number  = {7},
  pages   = {076301},
  year    = {2025},
  doi     = {10.1103/PhysRevLett.134.076301},
  eprint  = {arXiv:2308.01351},
  archivePrefix = {arXiv},
  primaryClass = {cond-mat.dis-nn}
}

@article{Boettcher2015,
  author  = {Igor Boettcher and Gregory Falkovich and Jan Schmalian},
  title   = {Critical Phenomena in Hyperbolic Space},
  journal = {Physical Review B},
  volume  = {92},
  pages   = {134423},
  year    = {2015},
  doi     = {10.1103/PhysRevB.92.134423}
}

@article{EversMirlin2008,
  author  = {Ferdinand Evers and Alexander D. Mirlin},
  title   = {Anderson transitions},
  journal = {Rev. Mod. Phys.},
  volume  = {80},
  pages   = {1355--1417},
  year    = {2008},
  doi     = {10.1103/RevModPhys.80.1355}
}

@book{Efetov1997,
  title={Supersymmetry in Disorder and Chaos},
  author={Efetov, Konstantin B.},
  year={1997},
  publisher={Cambridge University Press},
  address={Cambridge, UK}
}

@book{Altland2023,
	title = {Condensed {Matter} {Field} {Theory}},
	isbn = {978-1-108-49460-1},
	publisher = {Cambridge University Press},
	author = {Altland, Alexander and Simons, Ben},
	year = {2023},
}

@article{Jahn2019MatchgateHolography,
  author = {Jahn, Alexander and Gluza, Marek and Pastawski, Fernando and Eisert, Jens},
  title = {Holography and criticality in matchgate tensor networks},
  journal = {Science Advances},
  volume = {5},
  number = {8},
  pages = {eaaw0092},
  year = {2019},
  doi = {10.1126/sciadv.aaw0092},
  eprint = {1711.03109},
  archivePrefix = {arXiv},
  primaryClass = {quant-ph}
}

@article{Jahn2022Boundary,
  author = {Jahn, Alexander and Gluza, Marek and Verhoeven, Charlotte and Singh, Sukhbinder and Eisert, Jens},
  title = {Boundary theories of critical matchgate tensor networks},
  journal = {Journal of High Energy Physics},
  volume = {2022},
  number = {4},
  pages = {111},
  year = {2022},
  doi = {10.1007/JHEP04(2022)111},
  eprint = {2110.02972},
  archivePrefix = {arXiv}
}

@inproceedings{Meyer2003DiscreteDG,
  author    = {Mark Meyer and Mathieu Desbrun and Peter Schr{\"o}der and Alan H. Barr},
  title     = {Discrete Differential-Geometry Operators for Triangulated 2-Manifolds},
  booktitle = {Visualization and Mathematics III},
  editor    = {Hans-Christian Hege and Konrad Polthier},
  publisher = {Springer Berlin Heidelberg},
  address   = {Berlin, Heidelberg},
  year      = {2003},
  pages     = {35--57},
  isbn      = {978-3-662-05105-4},
  doi       = {10.1007/978-3-662-05105-4_2}
}

@article{OKeefe1998CoordinationSF,
  title={Coordination sequences for hyperbolic tilings},
  author={M. O'Keefe and Keeffe},
  journal={Zeitschrift Fur Kristallographie},
  year={1998},
  volume={213},
  pages={135-140},
  url={https://api.semanticscholar.org/CorpusID:96567465}
}

@article{BauesPeyerimhoff2001,
  author = {Baues, Oliver and Peyerimhoff, Norbert},
  title = {Curvature and Geometry of Tessellating Plane Graphs},
  journal = {Discrete \& Computational Geometry},
  volume = {25},
  number = {1},
  pages = {141--159},
  year = {2001}
}

@article{Keller2011,
  author = {Keller, Matthias},
  title = {Curvature, Geometry and Spectral Properties of Planar Graphs},
  journal = {Discrete \& Computational Geometry},
  volume = {46},
  pages = {500--525},
  year = {2011},
  note = {See Section 2.5 for volume growth bounds}
}

@misc{repo,
  author       = {Altland, Alexander and Micklitz, Tobias and Sharma, Devasheesh and Usoltcev, Maksimilian and Wille, Carolin},
  title        = {Supporting material for ``Theory of Anderson localization on the hyperbolic plane''},
  year         = {2026},
  publisher    = {Zenodo},
  version      = {1.0.0},
  doi          = {10.5281/zenodo.19728354},
  url          = {https://doi.org/10.5281/zenodo.19728354},
  note         = {Zenodo repository}
}


\newpage

\begin{appendix}

\begin{widetext}

\section{Laplace Operator on the hyperbolic plane}

Writing $\rho=r/L$ for the polar coordinates introduced in the main text, the Laplace operator
splits as $\Delta=L^{-2}(\partial_\rho^2+\coth\rho\,\partial_\rho
+\sinh^{-2}\rho\,\partial_\phi^2)$ with the radial
part $\Delta_\rho=L^{-2}(\partial_\rho^2+\coth\rho\,\partial_\rho)$. The eigenvalues
$L^{-2}\nu_\lambda\equiv -L^{-2}(\lambda^2+1/4)$ and
eigenfunctions $\phi_{m,\lambda}(\rho,\phi)=P^{|m|}_{-1/2+i \lambda}(\cosh \rho)\exp(i m
\phi )$ are known in closed form in terms of Legendre functions. However,
here, we will not need their detailed form, all that matters is
that they form a complete set allowing to expand functions on the hyperbolic
plane as 
\begin{align*}
    f(x)=\sum_{m\in \mathbb{Z}}\int d\mu_\lambda f_{m,\lambda}\phi_{m,\lambda}(x)  ,\qquad
 f_{m,\lambda}=L^{-2}\int d^2x\sqrt{g} \, \overline{\phi_{m,\lambda}}(x)f(x),
\end{align*}
where the integration over the Plancherel measure $d\mu_\lambda=\frac{1}{2\pi}  \lambda \tanh( \pi
\lambda)d \lambda$ extends over the full real axis $\lambda\in \mathbb{R}$. We also
note the `addition theorem' 
\begin{align*}
    \sum_{m\in \mathbb{Z}} \bar \phi_{m,\lambda}(x) \phi_{m,\lambda}(x)=1,
\end{align*} 
and the completeness relation
\begin{align*}
    \left\langle f,F(\Delta) g \right\rangle =\int d^2 x \sqrt{g}\, f(x) F(\Delta) g(x)=L^2\sum_m\int d \mu_\lambda 
    \bar f_{m,\lambda}F( L^{-2}\nu_\lambda) g_{m,\lambda}. 
\end{align*}   

Conceptually $\lambda\sim k L$ plays a role analogous to that of a dimensionless scalar  momentum
$k$, scaled in units of the curvature radius.  Furthering
this analogy, we  decompose functions as $f=f_\textrm{s}+f_\textrm{f}$ into a
`slow' and a `fast' component, where $f_\mathrm{s}$ is defined as the
contribution of the interval $|\lambda|L^{-1}\in [0,\Lambda e^{-\ell} ]$ to the mode
decomposition, and $f_\mathrm{f}$ as that of the interval $|\lambda|L^{-1}\in
(\Lambda e^{-\ell},\Lambda)$,
where $\Lambda$ is  a dimensionful high momentum cutoff, and $0< e^{-\ell}<1$ a scaling parameter.

\section{Renormalization group analysis}

Turning to our field theory, we represent our fields $T=\exp(W)$ in terms of
generators $W=W(x)$. Since these matrices live in the Lie algebra of the
symplectic group --- a linear space --- they can be decomposed as
$W=W_\textrm{s}+W_\textrm{f}$ into slow and fast components. With $T_\textrm{x}=\exp(W_\textrm{x})$, $\textrm{x}=\textrm{s,f}$, we
obtain the decomposition $T=T_\textrm{s} T_\textrm{f}$. We note that the symplectic
symmetry $T^{-1}=\sigma_2 T^T\sigma_2$ implies the constraint $W=-\sigma_2 W^T \sigma_2$.

This representation is the starting point for the perturbative integration over
$W_\textrm{f}$, following a textbook~\cite{Efetov1997,Altland2023} protocol:
substituting it into the action Eq.~\eqref{eq:nlsm}, we obtain the action $S=S_\textrm{f}
+ S_\textrm{sf}$, as the sum of two pieces, where (writing $W_\textrm{f}=W$ for
simplicity)
\begin{align*}
    S_\textrm{f}[W]=-\frac{\sigma}{16} 
\int d^2x\sqrt{g}\, \str \left( Q_\textrm{f} \Delta Q_\textrm{f} \right)\approx
- \frac{\sigma}{4} \int d^2 x\sqrt{g}\,\str(W \Delta W)=
 \frac{\sigma}{4}\sum_m \int d \mu_\lambda \str(\overline{ W_{m,\lambda}}  W_{m,\lambda}) \nu_\lambda
\end{align*} 
contains the expansion of the fast action to second order in Laplace eigenmodes with
$W_{m,\lambda}$ the matrix of Fourier coefficients of $W(x)$.
The remainder of the action, coupling slow and fast modes, is given by 
\begin{align}
    \label{eq:S0PriorRenormalization}
    S_\textrm{sf}[W,\Phi]\approx -\frac{\sigma}{4}\int d^2 x \sqrt{g} g^{\alpha \beta} \str(\Phi_\alpha \Phi_\beta 
    - [\Phi_\alpha,W][\Phi_\beta,W] ),  
\end{align}
where $\Phi_\alpha = T \partial_\alpha T^{-1}$, with $T\equiv T_\textrm{s}$. Doing the Gaussian integral over
$W$, in a one-loop approximation, $e^{S_\textrm{sf}[W,\Phi]}\to  e^{\left\langle
S_\textrm{sf}[W,\Phi] \right\rangle_W }  $, where $\left\langle \dots \right\rangle_W $ is
the Gaussian $W$-integral with weight $S_\textrm{f}[W]$, the second term turns into
$\str(\Phi_\alpha \Phi_\beta) C$, where the constant $C=\left\langle
|W^{ab}(x)|^2 \right\rangle_W $, is the average of two $W$-matrix elements at
coinciding points.

Up to this point, the analysis has been identical to that of the standard
two-dimensional nonlinear $\sigma$-model. The place
where the geometry of the hyperbolic plane enters is the `diagonal element of
the inverse Laplacian', encoded in the constant $C$. To see how, we turn to an
expansion in eigenfunctions, 
\begin{align*}
    C=\left\langle
|W^{ab}(x)|^2 \right\rangle_W=\sum_{m,m'}\int_\mathrm{f}  d \mu_\lambda d \mu_{\lambda'} 
\left\langle \overline W^{ab}_{m,\lambda}   
W^{ab}_{m',\lambda'} \right\rangle_W   \overline \phi_{m, \lambda}(x)
\phi_{m', \lambda'}(x)=\frac{2}{ \sigma }\int_\mathrm{f} 
 d\mu_\lambda \frac{1}{\nu_\lambda},
\end{align*} 
where we used the addition theorem, and $\int_\mathrm{f}d\mu_\lambda $ is
shorthand notation for the integral over the domain of fast $\lambda$-values.
(Physically, the collapse of the $m$-sum over angular harmonic reflects the
isotropy of the Laplace operator.) With this result, we obtain the
renormalization of the conductivity  
as in a single RG step as Eq.~\eqref{eq:SigmaOfb},
\begin{align*}
    \sigma\to \sigma(\ell)=\sigma_0 - \frac{1}{2\pi} \int^{\Lambda L}_{\Lambda  L e^{-\ell}} 
    \frac{d\lambda \lambda \tanh(\pi \lambda)}{\lambda^2+\frac{1}{4}}.
\end{align*}

After the execution of the RG step, we rescale variables to restore the original
cutoff from the lowered value $\Lambda e^{-\ell}$ back to $\Lambda$. This step is
not entirely innocent, as we need to decide what to do with the scale-dependent
curvature. Substituting the one-step updated $\sigma(\ell)$ as a prefactor into
Eq.~\eqref{eq:S0PriorRenormalization} and switching back to the invariant
representation, we obtain the action
\begin{align*}
    S_0[Q]=-\frac{\sigma(\ell)}{16} \int_{\Lambda e^{-\ell}} d^2 x \sqrt{g(x/L)} 
    g(x/L)^{\alpha \beta} \str(\partial_\alpha Q \partial_\beta Q),
\end{align*}
where the subscript of the integral indicates that the spatial integration is
over slow fields with a minimal resolution scale of $\Lambda e^{-\ell}$, and we
indicate the dependence of the metric on the dimensionless ratio $x/L$. In order
to restore the original cutoff, we rescale the spatial coordinates as $x\to x
e^\ell$. This brings the action back to the original form, with, however, an
updated curvature radius $L(\ell)=e^{-\ell}L$: after the RG step, the curvature
radius has become smaller in relation to the updated short distance cutoff.

The presence of the scale dependent length, $L(\ell)$ entering the RG flow
implies that we are outside the standard `one-parameter scaling' paradigm of
Anderson localization. To describe this flow in terms of RG equations, we
introduce the dimensionless scaling variable $u(\ell)=\Lambda L e^{-\ell}$
and differentiate $\frac{d\sigma(\ell)}{d\ell}\big|_{L=L(\ell)}$ with respect to
$\ell$, at the rescaled parameter $L=L(\ell)$. This leads to the two coupled
flow equations in Eq.~\eqref{eq:TwoParameterFlow}.

Finally, note that the rescaling of $L$ could have been implemented right away
on the basis of dimensional analysis. It is worth mentioning that older work on
renormalization of field theories on curved
spaces~\cite{JackOsborn1990,Osborn1991,Vassilevich2003} avoids this rescaling,
and in this way effectively changes the geometry. This intrusion is then
compensated for by a nontrivial update of the coefficient $\sigma(\ell)\to
\sigma_\mathrm{geo}(\ell) $, called `projection'. The motivation for this
alternative scheme has to do with the  status of the metric in quantum gravity,
as an independent background field or dynamical variable.


\end{widetext}

\section{Lattice localization}

In this section, we provide details concerning the localization transition on discretizations of the hyperbolic plane, specifically the PD discretization. The analysis is based on generalizing the well-known mechanism of Bethe lattice localization to the presence of loops and fluctuating weights. 
As a warmup, we begin with a review of the localization
transition on the Bethe lattice. 

\subsection{Bethe lattice localization}

Consider a Bethe lattice, fix a site, $0$, and denote by $\Psi(Q)=\int_Q
DQ\exp(-S[Q])$ the path-integral over 
the subtree connected to this site with the boundary
condition $Q_0=Q$. With, $L_{w}(Q,Q')\equiv \exp(-w\, \str(QQ'))$, we
may represent this path integral through the recursion relation
\begin{align}
    \label{eq:BetheRecursion}
    \Psi_i(Q)=\int
dQ' \,\prod_j L_{w_{ij}}(Q,Q') \Psi_j(Q'),
\end{align} 
i.e. the contribution of $K=q-1$ daughter sites, connected to $0$ via a single
stem, where the early theory of Bethe lattice localization considered the case
of uniform weights $w_{ij}=w$. In the localized regime, $w\ll 1$, the integrals
over the non-compact $Q$-matrix manifolds are dominated by non-compact
`radial' coordinates, parametrizing the unbounded integration volume of the
$Q$-manifold. While our model manifold  $\textrm{OSp}(2|2,4)$
 has two of these, we here consider a simplified  variant with just one
radial coordinate, $t$, required to span the smaller  
$\textrm{U}(1|1,2)$ of a model neglecting the time-reversal symmetry. (While time-reversal symmetry is essential to weak localization, the principles of strong disorder localization do not depend on it~\cite{Efetov1997}.) 
With the ansatz
$\Psi(Q)\to \Psi(t)$, the integration over all other matrix variables then
leads to a reduction $L_w(Q,Q')\to L_w(t-t')$, with
\begin{align}
    \label{eq:KernelLt}
    L_w(t)\approx \left( \frac{w}{2\pi} \right)^{1/2} \exp\left(-w \cosh t+\frac{t}{2}\right). 
\end{align}   
 Deep  in the localized regime, $w\to 0$, the integral decouples,
$L(Q,Q')\to 1$, and $\Psi(Q)=1$ solves the recursion relation (formally as a
consequence of supersymmetry). We will identify the localization transition with
the threshold $w=w^\ast$ where this solution becomes unstable. 

To identify this instability, we expand $\Psi(Q)\to \Psi(t)=1+ \Phi(t)$, and
consider the linearized equation $\Phi(t)= K (L\ast\Phi)(t)$, with the
convolution $(f\ast g)(t)=\int ds\,f(t-s)g(s)$. Referring to Refs.~\cite{Efetov1997,ZIRNBAUER1986375} 
for a detailed discussion including all
convergence generating factors, the stability of the solution hinges upon the
existence of a unit eigenvalue, $\kappa=1$, of the linear kernel $KL$. Thanks to
its translational invariance, $L(t,t')=L(t-t')$, eigenfunctions assume an
exponential form, and a  variation of the eigenvalue equation
$K\int dt' L(t-t') \exp(-\theta(t-t'))=\kappa $ shows that the minimal eigenvalue
is obtained for $\theta=1/2$. Plugging this value back into the integral, a
straightforward estimate yields the criterion for a unit eigenvalue $\kappa^\ast
=1$ as
\begin{align}
    \label{eq:UnitEigenvalueHomogeneous}
    1\approx K\left( \frac{w^\ast}{2\pi} \right)^{1/2} \log \left( \frac{2}{w^\ast} \right),
\end{align}
whose solution defines the critical coupling as quoted in the main text.

\subsection{Poisson-Delaunay localization}

\begin{figure}
    \centering
    \vspace{.1cm}
    \includegraphics[width=3.5cm]{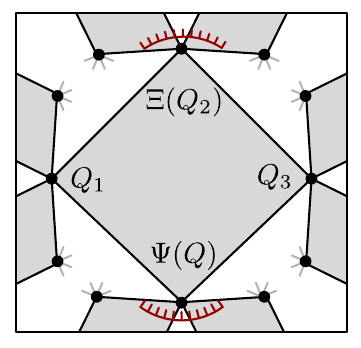}
    \vspace{-.2cm}
    \caption{Two cell generations of a cactus with $q=3$ $(p=4)$-gons meeting at
    each vertex.  
 }    
\label{fig:figure4}
\end{figure}

As mentioned in the main text, the generalization of the theory to the PD random
graph requires accounting for loops and randomly distributed weights. 

\paragraph{Loops:} Referring to Fig.~\ref{fig:figure4} for a graph comprising
polygon-like loops ($p=4$ in the figure) meeting at $q$-polygon vertices (the
analysis will show that the detailed values $(p,q)$ do not matter and may even
fluctuate across the graph) we first identify a recursion relation replacing
Eq.~\eqref{eq:BetheRecursion}: Singling out a vertex with field
value $Q$, we consider the contribution $\Psi(Q)$ of one specific plaquette
to the partition sum, e.g. the integral over all fields sitting at vertices
north of the one at the bottom of the figure. Integrating over the fields
$Q_1,\dots,Q^{p-1}$, at the other vertices inside that plaquette, we obtain the
first of two equations, 
\begin{align*}
    \Psi(Q)=\int_{ Q_1 \dots Q_{p-1}}\hspace{-1cm}L(Q,Q_1)
    \dots L(Q_{p-1},Q)\Xi(Q_1) \dots \Xi(Q_{p-1}).
\end{align*} 
Here, the weights $\Xi(Q_i)$ are the contributions of all plaquettes
excluding the one we are considering to the vertex partition sums $Z(Q_i)$. By
construction, this is $\Xi(Q_i)=\Psi^K(Q_i)$, which closes the recursion.

Turning to its solution, we again linearize, and focus on the leading noncompact
variable $\Psi(Q)\to 1+\Phi(t)$. The insertion of this approximation into the
recursion relation then leads to expressions of the structure $q\times L^{\ast l}\Phi
L^{\ast p-l+1}$, i.e. repeated convolutions of the kernel $L$  sandwiching a
single distortion $\Phi(t)$ under the integral. While the analytical computation
of the convoluted kernel $K^{(l)}(Q,Q')\equiv (L^{\ast l})(Q,Q')$ is a
complicated affair (See Ref.~\cite{repo} for details),
the result can be understood heuristically: in the limit $w\lesssim 1$, multiple integration
over intermediate site-variables effectively describes the consecutive tunneling
through $l$ tunnel contacts. We also know that the integration
must produce an invariant function of the argument matrices, $Q$ and $Q'$. While the
actual result looks more complicated, its scaling in the limit $w\lesssim 1$ is
described by  $K^{(l)}(Q,Q') \approx \exp(-w^l
\str(QQ'))$. Individual terms in the sum will then scale
as $\sim \exp( -(w^l+w^{p-l+1})  \str(Q,Q'))\Phi(Q')$, indicating that the
dominant contributions come from the maximally asymmetric configurations $l=1$
and $l=p$. Physically, these represent 
quantum transport through the shortest possible links, e.g., $Q\to Q_1$ and
$Q\to Q_3$ in  Fig.~\ref{fig:figure4}. The coupling
along these receives a mild upward correction $w\to w +w^p$ due to the
inclusion of the $p$-step complementary segment of the loop e.g. $Q\to Q_3 \to Q_2 \to
Q_1$. This correction implies that slightly larger disorder (smaller $w$) is
required to induce localization on the loopy lattice. Otherwise, however, the
physics of the system remains similar to that of a Bethe lattice of
coordination number $q$.

\paragraph{Random couplings:} Consider a situation, where the couplings $w_{ij}$ exhibit $\mathcal{O}(1)$
fluctuations reflecting the geometric fluctuations of the randomly drawn
lattice. In this case, the linearization of the recursion
Eq.~\eqref{eq:BetheRecursion} gets us to the linear integral equation
$\Phi_i=\sum_j (L_{w_{ij}}\ast \Phi_j)$, and the scalar equation
\eqref{eq:UnitEigenvalueHomogeneous} generalizes to
\begin{align*}
   \Phi_i = \sum_j\kappa_{ij} \Phi_j,\qquad \kappa_{ij}= \left( \frac{w_{ij}}{2 \pi} \right)^{1/2}\ln \left( \frac{2}{w_{ij}} \right).    
\end{align*} 
For a PD graph, the coefficients $w_{ij}$ exhibit $\mathcal{O}(1)$ fluctuations
around their mean value $w$, and in view of the large number of sites in the
lattice it seems reasonable that the equation affords a solution if the
homogeneous criterion Eq.~\eqref{eq:UnitEigenvalueHomogeneous} holds. In other
words, we take the structure of the equation as indication that the presence of $\mathcal{O}(1)$ bond strength
fluctuations on a random lattice does not significantly affect its localization
properties.

\section{Lattice discretizations of \texorpdfstring{$\mathbb{H}^2$}{H^2}}

We here provide an overview of $\mathbb{H}^2$ discretizations, focusing on the three
criteria discussed  in the text: 
First, the lattice discretizations need to reproduce the exponential growth of metric area, characteristic for the hyperbolic plane where the boundary of a ball with radius $r$ grows as $|\partial B_r| \sim e^{r/L}$. For  discretizations it is natural to replace the notion of radius $r$ by a discrete layer index $i$ (corresponding to the $i$-th generation in the Bethe or Husimi cactus tree, or the $i$-th ring of polygons in a $\{p,q\}$-tessellation) and define the area growth rate $g$ via the number $s_i$ of vertices in each layer as $s_i \sim g^i$ such that $\log g \sim L^{-1} $ sets an effective  curvature scale.
%
Second, a 
graph's \emph{loop concentration}~\cite{Hashimoto1989,Terras2010}, is defined through the ratio $c_n=C_n/N_n$ of closed ($C_n$), vs arbitrary ($N_n$) non-backtracking paths of length $n$. Third, the level of agreement between lattice and continuum
Laplacian is determined by comparison of their spectra.

\begin{figure}[t!]
    \centering
    \vspace{.1cm}
    \includegraphics[width=8.5cm]{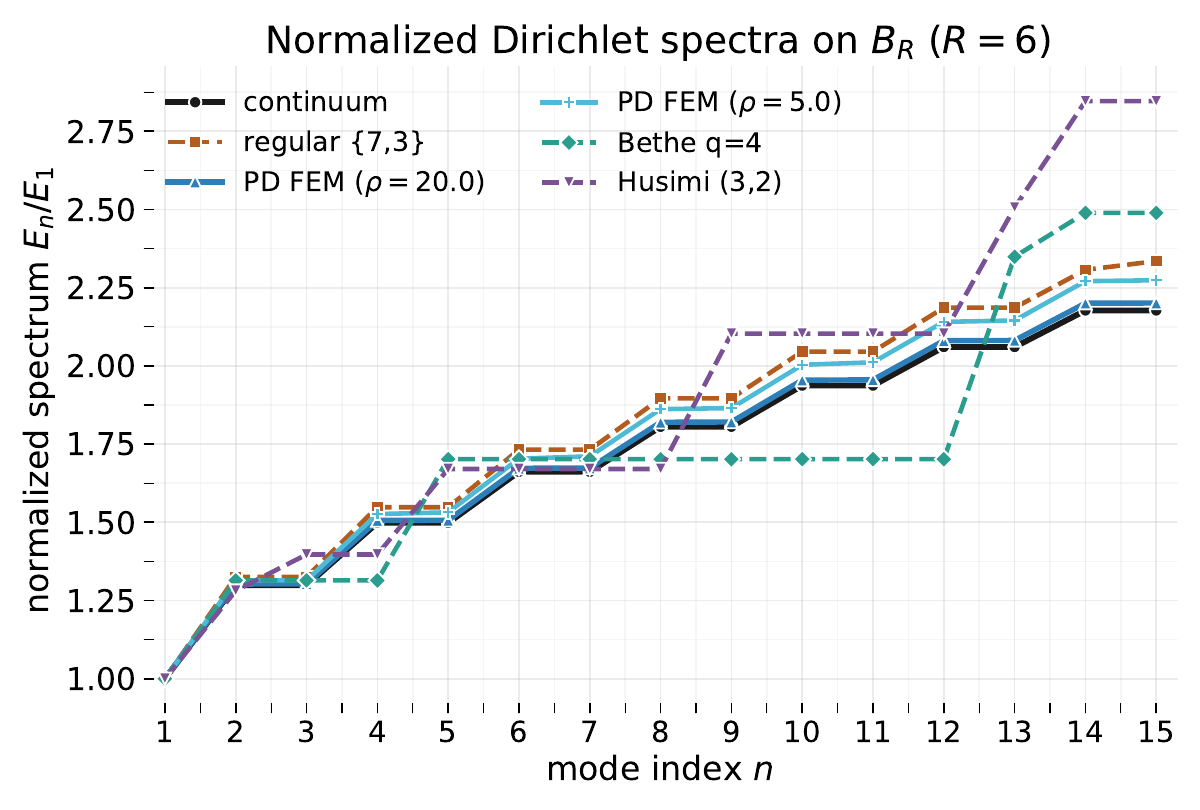}
    \vspace{-.2cm}
    \caption{Spectra of different graph discretizations of a truncation of
    $\mathbb{H}^2$ at a finite radius $R$ in units of curvature radius $L$, compared to the  continuum theory
    (black) of the same geometry with Dirichlet boundary conditions. The curves
    are scaled manually to a coinciding lowest mode value. (The
    latter is finite, and determined by the effective curvature whose detailed
    comparison between different discretizations is not straightforward.) We
    notice strong deviations between the continuum and both Bethe, and Husimi
    cactus discretizations. The regular $\{7,3\}$ tessellation shows much better
    agreement, with, however, increasing deviations beyond the lowest lying modes.
    As expected, the PD discretization provides the best proxy of the
    continuum spectrum, with its accuracy increasing as the mesh density $\rho$ grows.  }    
\label{fig:figure5}
\end{figure}

\paragraph{Poisson-Delaunay triangulation \cite{BenjaminiPaquettePfeffer2014}:}
A discretization in terms of random points in $\mathbb{H}^2$ drawn from a uniform
measure of density $\rho$. The Delaunay triangulation of these points produces
a statistically homogeneous graph with average coordination approaching $\langle q\rangle=6$~\cite{NielsenNock2009} in the high-density,
locally flat limit. The connectivity $w_{ij}$ between nearest
neighbor points is determined by the local triangle geometry via a
standard finite-element (cotangent) construction~\cite{Meyer2003DiscreteDG},
resulting in a lattice Laplacian whose continuum limit reproduces the Laplace
operator by design (see Fig.~\ref{fig:figure5}).
The PD discretization features strong fluctuations of the edge weights, which encode the geodesic distances. It is therefore most natural to measure the volume growth not by a combinatorial distance measure, but with respect to the geodesic distance $r$. The growth rate then matches that of the continuum, $ \log g \sim L^{-1}$. 
Similarly, the bare combinatorial loop content of the PD triangulation is agnostic to the weights and is not a physically meaningful quantity. However, return probabilities that factor in the length of loops $l$ via the respective edge weights show exponential decay as apparent from the absence of diffusion in a random walk (see Fig.~\ref{fig:figure1}). 
The spectrum of the graph Laplacian approaches that of the continuum Laplacian in the limit of increasing the sampling rate/mesh density $\rho$ as shown in Fig.~\ref{fig:figure5}.
%

\paragraph{Bethe lattice:}
A $q$-regular tree. The growth rate is equal to the branching number $g=K=q-1$ and there are no loops, $C_n=c_n=0$. There is no limit in which the spectrum of the graph Laplacian approaches that of the continuum Laplacian.

\paragraph{Husimi cactus graph:}
A generalization of the Bethe lattice in which $q$ $p$-gons meet at each
vertex. The growth rate remains proportional to the (polygon)-branching number $K_q=(q-1)$, but is augmented by the vertices per polygon, $g=(q-1)(p-1)$. The probability of a non-backtracking path to be closed decays exponentially with the path length at a rate that is largely determined by the branching number $K_q$. The largest loop density is that of a single polygon loop, $n=p$, which occurs with probability $(2q-1)^{-p}$. As for the Bethe lattice, the spectrum of the graph Laplacian does not match that of the continuum.

%

\paragraph{Regular $\{p,q\}$-polygon tessellations:}
A popular choice for
$\mathbb{H}^2_{L}$ discretizations, realized in terms of $p$-gons meeting at
$q$-valent vertices.
If $a$ denotes the edge length, the curvature
radius is fixed by
$\cosh\!\left(\frac{a}{2L}\right)=\frac{\cos(\pi/p)}{\sin(\pi/q)}.$
For example, $L/a\simeq 1.76$ for the $\{7,3\}$
tessellation shown in Fig.~\ref{fig:figure3}. 
At fixed $q=3$, increasing $p$ moves the
tessellation away from the Euclidean $\{6,3\}$ limit and decreases
$L/a$; for instance, $\{20,3\}$ has $L/a\simeq 0.95$.
The growth rates lie in the window 
$ 1 + \frac{(p-2)(q-2) - 4}{p-1}\le g  \le q-1$ \cite{Keller2011,BauesPeyerimhoff2001} and can be calculated explicitly \cite{OKeefe1998CoordinationSF} albeit not necessarily in closed form. The loop concentration decays exponentially with the loop length with a decay rate proportional to $1/L$. The (normalized) Laplacian spectra agree well with the continuum spectra for low eigenvalues, but differ increasingly higher up in the spectrum.




\section{Conductance calculation}
\label{sec:ConductanceCalculation}

We here outline how the scale-dependent resistance $R(r)$ of the hyperbolic
plane is obtained from  the effective action
Eq.~\eqref{eq:nlsm}. Considering the two-terminal setup mentioned in the text,  its
metallic electrodes introduce  boundary conditions
$Q(0)=Q(d)=\tau_3$ \cite{Efetov1997}. We introduce a voltage bias by 
coupling  the action to a response vector potential in
radial direction $\partial_rQ\to \partial_r Q- [A_r,Q]$, where $A_r\equiv
(a\tau^++ \bar a \tau^-) X$, $\tau^\pm = \frac{1}{2}(\tau_1 \pm i \tau_2)$,  and
$X=\sigma_3 \otimes P^1$ contains a Pauli matrix $\sigma_3$ in time-reversal
space, and a projector $P^1$ onto the first auxiliary channel. Physically, this
generalization exposes the system to a constant radial field gradient. Formally,
it serves as a source field, from which the conductance is obtained by two-fold
differentiation,  $g=d^{-2}d^2_{a\bar a}\big|_{a=0}\int DQ\exp(-S[Q,A])$~\cite{Efetov1997}.

In
the semiclassical limit of large conductivity, we evaluate this response functional on
a background field configuration $Q=T\tau_3 T^{-1}$, with smoothly varying
$T=T(r)$. (The exclusion of strongly fluctuating modes on top of a smoothly
varying one effectively means that we have followed the RG flow to a scale
$\Lambda^{-1}\exp(\ell)\sim d$ but no further.)  We
choose this rotation matrix to swap the spatially constant radial field
represented by $A_r$ by a spatially varying one, $\partial_r V(r)$. This
replacement is achieved by the ansatz $T\equiv
\exp((V(r)-r)A_r)$, whose substitution into the covariant derivative effectively
changes $A_r \to \partial_r V(r) (a\tau^++ \bar a \tau^-) X $. In order not to
change the metallic boundary conditions, we require $V(0)=0$ and $V(d)=d$, i.e.
$T(0)=T(d)=\mathrm{id} $. With  $\partial_r \tau_3=0$, the straightforward
substitution of the commutators $[A_r,\tau_3]$ into the action followed by the
execution of the matrix trace yields the action $S[Q,A]= \sigma(d)a \bar a\int \sqrt{g}
(\partial_r V(r))^2 $. Differentiating with respect to the source parameters and
doing the trivial angular integral, we obtain  the conductance 
\begin{align}
    \label{eq:GFromVFunctional}
 g(d)=  \frac{\sigma(d)}{d^2}  \int_0^d dr \, C(r) (\partial_r V(r))^2\equiv S[V],
\end{align}
 where $C(r)=2\pi
\sqrt{g(r)}=2\pi L \sinh(r/L)$.  The physical value
of this functional follows from its evaluation on a stationary voltage profile,
$\delta_V S[V]=0$. Doing the derivative, we obtain  $\partial_r V(r)=c/C(r) $,
where the constant  is fixed by our boundary condition as $d =\int_0^d
dr\partial_r V =c\times  \int_\epsilon^d dr \,C(r)^{-1} $. Substituting these results
back into Eq.~\eqref{eq:GFromVFunctional}, we obtain 
\begin{align*}
    g(d)= \sigma(d)\left( \int_\epsilon^d dr \, C(r)^{-1} \right)^{-1},
\end{align*} 
where we regularized zero by a small $\epsilon>0$.
The elementary integral then leads to the final result for the resistance $R=g^{-1}$ in Eq.~\eqref{eq:ConductanceFromSigma}.

\end{appendix}


\end{document}